# Technical Evaluation of HoloLens for Multimedia

## A First Look

**Yang Liu**
*University of Ottawa*

**Haiwei Dong**
*University of Ottawa*

**Longyu Zhang**
*University of Ottawa*

**Abdulmotaleb El Saddik**
*University of Ottawa*

Augmented-reality (AR) technology has been developing rapidly for decades. A recently released cutting-edge AR device, Microsoft HoloLens, has attracted considerable attention with its advanced capabilities. In this paper, we report the design and execution of a series of experiments to quantitatively evaluate HoloLens' performance in head localization, real environment reconstruction, spatial mapping, hologram visualization, and speech recognition. The results show that HoloLens is able to estimate head posture more correctly at low movement speeds, reconstruct the environment most precisely for a flat surface under bright conditions, anchor augmented contents at desired locations most accurately at distances of 1.5 m and 2.5 m, display objects with an average size error of 6.64%, and recognize speech commands with correctness rates of 74.47% and 66.87% for user-defined and system-defined commands, respectively. Discussions are also provided to further explain our work and the limitations of the experiments.

Visualizing real and virtual contents simultaneously is of great importance in the field of augmented reality (AR). Significant progress has been made in this area, and many AR devices have been developed. For example, Google Glass and Epson Moverio BT-300 both use a pair of eyeglasses to display augmented contents superimposed on the real surrounding environment.[1-2] The recently released cutting- edge AR device HoloLens, developed by Microsoft, differs from most other such devices in that it itself is a complete AR system, running the Windows 10 operating system (OS) and containing a central processing unit (CPU), a custom-designed holographic processing unit (HPU), various types of sensors, see-through optical lenses with a holographic projector, and so forth (www.theverge.com), as show in Figure 1.

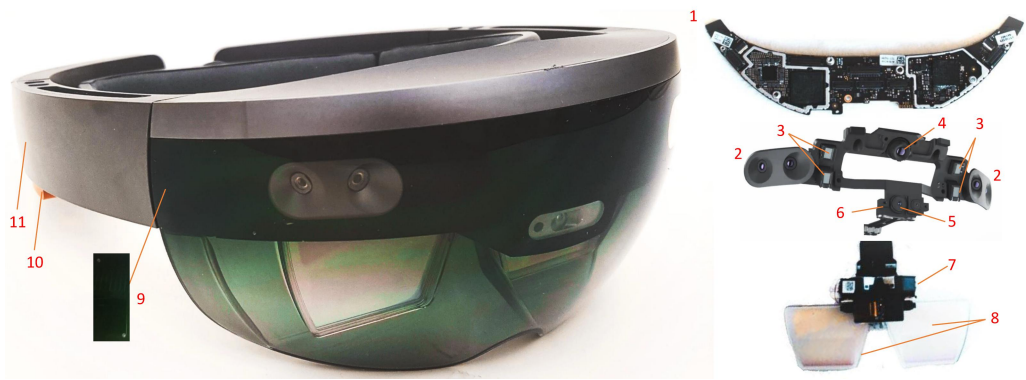

Figure 1. HoloLens hardware components. 1) Processing units. 2) Environment understanding cameras. 3) Infrared laser projector. 4) Depth camera. 5) HD video camera. 6) Ambient light sensor. 7) Holographic projector. 8) See-through optical waveguide lenses. 9) Microphones. 10) Built-in speaker. 11) Battery

The advanced performance of HoloLens has attracted considerable attention. Avila et al. presented the basic capabilities of HoloLens;[3] Zhang et al. utilized HoloLens to visualize and interact with smart city data;[4] and Lu et al. illustrated several examples of the use of HoloLens for immersive analysis and discussed the new opportunities and challenges the system presents for visualization and visual analytics.[5]

Since most research on HoloLens and its applications is highly dependent on the device itself, in this study, we conducted a series of technical evaluations of HoloLens' performance, based on its operational mechanism, to explore its capabilities and limitations. We begin by introducing HoloLens' five main functional components, and then explain our corresponding experimental designs and implementations for each component. Discussions are also provided to further introduce our work and address its limitations.

## HOLOLENS OPERATIONAL MECHANISM

As an independent AR system, HoloLens consists of several functional components. Based on the operational mechanism of HoloLens, we divide it into five main components: head pose estimation, real environment reconstruction, virtual environment processing, user perception, and user control.

The head posture of a user can be determined from the position and orientation of the HoloLens system, as estimated by means of the inertial measurement unit (IMU) and the iterative closest point (ICP) algorithm; the success of this posture determination substantially influences the tracking accuracy of the system, according to Zhang's paper.[6] The real environment can be reconstructed as a 3D model by means of the depth camera, the environmental understanding cameras, and the KinectFusion algorithm, the last of which was originally developed for 3D reconstruction using the Kinect depth camera.[7] Then, the virtual environment component receives information from the head posture, real environment, and user control components to augment specific locations in the real environment with computer-generated holograms through suitable processing of the obtained data. User perception in HoloLens mainly focuses on providing visual information by means of holographic projection technology and offering a realistic audio experience by means of spatial sounds.[8] User control refers to the ability of the user to interact with the holograms through gaze, gesture, and speech commands, among other types.

# TECHNICAL EVALUATION

Based on the main functional components described above, we designed a series of experiments to technically evaluate the performance of each HoloLens component: a head localization experiment, to compare the head posture estimation results from HoloLens with a ground-truth record from OptiTrack; a real environment reconstruction experiment, to evaluate the differences between the reconstructed model and the real environment; a spatial mapping experiment, to measure the gap or overlap between an augmenting hologram and the target mapping surface; a hologram visualization experiment, to calculate the deviation between a visualized hologram and its corresponding real object; and a speech recognition experiment, to test the reliability of user control through voice commands.

Below is presented some general information regarding all experiments. More details will be given later in each subsection.

- All experiments were conducted in a closed 8 m×5 m room under its ambient lighting conditions, controlled by means of lights with adjustable intensities and angles. This room could also prevent echoes during the experiments because of its sound-absorbing walls.
- The HoloLens applications used in our experiments were all developed with Unity and Visual Studio 2015 using the C# programming language for compatibility with HoloLens' Windows 10 OS.
- A total of 20 students and researchers from the University of Ottawa participated in our experiments. Their average age was 25.89, with a standard deviation of 6.11. They were either native or fluent English speakers. Each participant was given 10 minutes to become familiar with HoloLens and then spent approximately 25 minutes performing all experiments.

## Head Localization

*1) Design:* The head localization experiment was designed to evaluate the accuracy and stability of HoloLens' head posture estimation. To explore the effect of drift of the IMU and ICP evaluation, as previously mentioned, participants wearing HoloLens were asked to test various conditions: moving or rotating the head at high or low speeds in the x, y, and z dimensions separately. The head localization performance was then evaluated in terms of Euclidean distances, which represent the distance between two points in a metric Euclidean space, by comparing the results from HoloLens with the ground truth from OptiTrack.

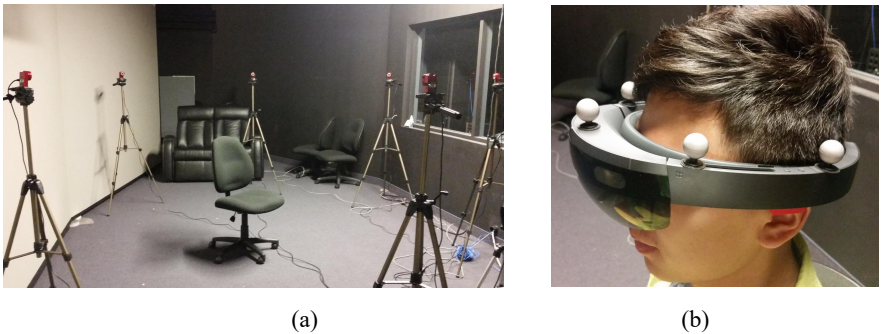

(a)            (b)

Figure 2. Experimental setup for the head localization experiment. (a) OptiTrack cameras placed in a circle with a 2 m radius. (b) User wearing HoloLens with IR markers. OptiTrack was used to record the ground truth for head localization based on the IR markers, and these records were then compared with the head localization estimation results from HoloLens.

*2) Materials:* In this experiment, the OptiTrack system was employed to record the ground truth for head localization. OptiTrack is an accurate marker-based motion capture system, with a distance error of ±0.3 mm and a rotational error of ±0.05 degrees. In this experiment, six OptiTrack sensors were placed in a circle with a 2m radius to track the positions of IR markers mounted on the HoloLens system. The OptiTrack system (Flex V100 camera with software Arena v1.7) was set up in the room as depicted in Figure 2a, and its tracking markers were attached to the HoloLens unit as shown in Figure 2b.

*3) Procedure:* In this experiment, first, the OptiTrack system was calibrated to accurately record the ground truth for head localization. Second, the HoloLens and OptiTrack systems were both activated to record tracking data. Third, the participant, wearing the HoloLens unit with markers, was asked to perform several actions in the center of the circle of cameras in a random order, including squatting quickly (fast movement), tilting the body slightly (slow movement), looking at the corner of the room quickly (fast rotation) and swinging the head gently (slow rotation). A laboratory technician was trained to demonstrate these actions during the experiment, and the participant was asked to simultaneously imitate the demonstrated action and speed.

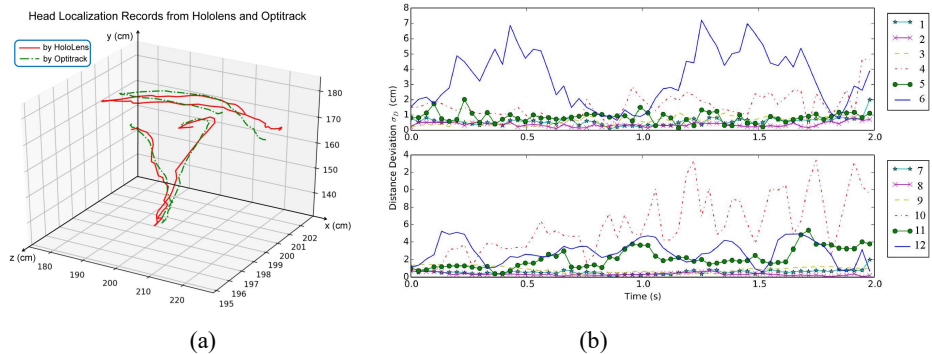

Figure 3. Results of the head localization experiment. (a) Example of tracking records from HoloLens and OptiTrack. (b) Distance deviations between the two records, where each group of three lines, 1-3, 4-6, 7-9, and 10-12, represents the distance deviations $\sigma_D$ in the x, y, and z dimensions for moving slowly, moving quickly, rotating slowly, and rotating quickly, respectively. The results show that HoloLens is able to estimate head posture more correctly at low movement speeds.

*4) Results:* An example of tracking records from HoloLens and OptiTrack is shown in Figure 3a, where the solid red lines represent the head localization records from HoloLens and the dotted green lines represent the records from OptiTrack. The distance deviations $\sigma_D$ between the two records are shown in Figure 3b. The upper subfigure shows the distance deviations $\sigma_D$ caused by head movement, and the lower subfigure shows the distance deviations $\sigma_D$ caused by head rotation. Since the distance deviations $\sigma_D$ under each condition were all measured in a 3D coordinate system, groups of three lines (lines 1- 3, 4-6, 7-9, and 10-12 in Figure 3b) are used to present the results in the x, y, and z dimensions for moving slowly, moving quickly, rotating slowly and rotating quickly, respectively. The average distance deviation values are 0.53, 1.63, 0.60, and 3.62 cm, with standard deviations of 0.03, 0.63, 0.02, and 0.47 cm, respectively. The record values from HoloLens and OptiTrack do not have significant difference ($F_{1,20}$ = 1.96, P = 0.52). At high speed, it is difficult for HoloLens to correct the head localization results. This can be seen from the fact that the average distance deviation $\sigma_D$ in the fast mode (both movement and rotation) is 2.63 cm, whereas it is 0.56 cm in the slow mode. The highest distance deviation $\sigma_D$ is 13.38 cm, caused by the head rotating quickly along the x axis.

## Real Environment Reconstruction

*1) Design:* The real environment reconstruction capability of HoloLens was evaluated by comparing a real environment with its reconstructed 3D model. Such a reconstructed model is usually influenced by the complexity of the real environment.[9] The performances of the sensors are also influenced by lighting conditions, surface textures, and the distances between HoloLens and the real objects. Therefore, in this experiment, the differences between a real environment and its reconstructed model were separately measured for different influencing factors: object shapes (flat, convex or concave angles), lighting conditions (bright or dark).

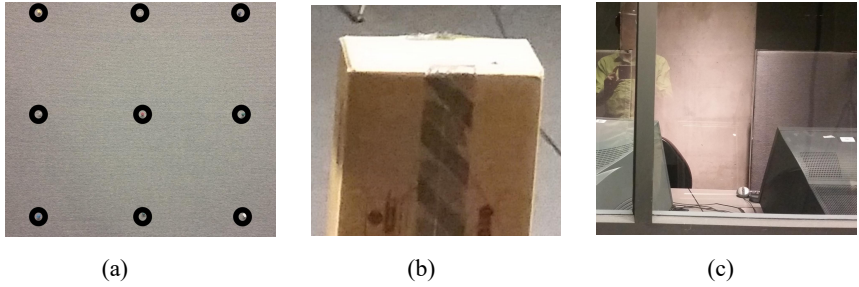

(a)            (b)            (c)

Figure 4. The marked objects used for testing the accuracy deviation $\sigma_A^R$ of the reconstructed models obtained from HoloLens. (a) Flat surface with markers. (b) Box placed at a convex angle. (c) Glass surface separating the participant wearing the HoloLens system from objects under bright light.

*2) Materials:* In this experiment, the reconstruction performance for object shapes was evaluated using a flat surface (Figure 4a) and a box with convex and concave angles (Figure 4b). The brightness of the environment was controlled using 10 brightness-adjustable incandescent lamps. The distance between the lamps and the object to be measured was 3.5 m. For the bright lighting condition, the output power of the lamps was set to 25 W, whereas their power was reduced to 5 W for the dark lighting condition. When the participants recorded markers through a glass pane (Figure 4c), the markers were placed under the bright lighting condition. In the measurement implementation, the gaze point is visualized as a red circle, and its corresponding position information is also displayed nearby.

*3) Procedure:* During the experiment, first, the participants were asked to walk around the room to reconstruct a 3D model of the room under the bright lighting condition. Second, the experimental data acquired by HoloLens were recorded. Third, the participants were asked to place the red circle of the gaze-tracking application on the marked locations to obtain the dimensions of the objects' reconstructed models under bright and dark lighting conditions or through a glass pane placed at a certain distance. For example, the participants were asked to place the red circle on the 9 markers on the flat surface (shown in Figure 4a) to obtain the lengths between the reconstructed markers. Similarly, the participants were asked to place the red circle on the two vertices of the box corresponding to each side of the box lying at the convex or concave angle to calculate the length of the box. To avoid interference between the different influencing factors, the sequence of the experiment was varied for different participants.

The reconstruction accuracy $\sigma_A^R$ was evaluated based on the difference between the real environment and its reconstructed model:

$$\sigma_A^R = \sum_{i \subset N} (\frac{|L - l_i|}{L \cdot N}) \qquad (1)$$

where $N$ represents the number of measurements, $L$ is the length of one edge of the real object, and $l_i$ is the corresponding edge length measured in the $i$-th measurement by HoloLens. $R$ is used to denote that this reconstruction accuracy refers to the real environment reconstruction performance. The values of $\sigma_A^R$ are greater than or equal to 0, where 0 means that the real object and its corresponding holographic model are identical and higher $\sigma_A^R$ values indicate an increasing difference between the object and the model.

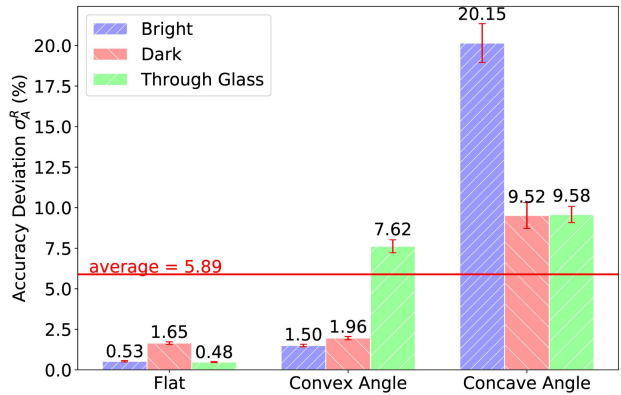

Figure 5. The reconstruction accuracy $\sigma_A^R$ for real environment reconstruction. The results show that HoloLens reconstructs the environment most precisely for a flat surface under bright conditions.

*4) Results:* The reconstruction accuracy $\sigma_A^R$ for the tested objects are shown in Figure 5, where the error bars represent the standard deviation for each condition, the blue and red bars represent the object size deviations captured under the bright and dark lighting conditions, respectively, and the green bar represents the reconstruction accuracy $\sigma_A^R$ for an object behind glass. The average $\sigma_A^R$ is 5.89%, and its standard deviation is 6.18%. The lowest $\sigma_A^R$ value is 0.53%, which is obtained when the object's surface is flat and it is under bright light, whereas the highest $\sigma_A^R$ value is 20.15%, which is obtained when the object is a box at a convex angle. This indicates that better environmental reconstruction results can usually be obtained with flat surfaces under bright lighting conditions compared with the reconstruction of uneven surfaces (convex or concave) in dark environments.

## Spatial Mapping

*1) Design:* The spatial mapping experiment, which investigated the anchoring of holograms to the real environment, was performed to test the virtual environment processing component. In our experiment, we measured the gap or overlap between a hologram and the target surface to which it was attached, and the spatial mapping performance was evaluated using a measure denoted by $\sigma_A^S$ (calculated in a manner similar to Equation 1).

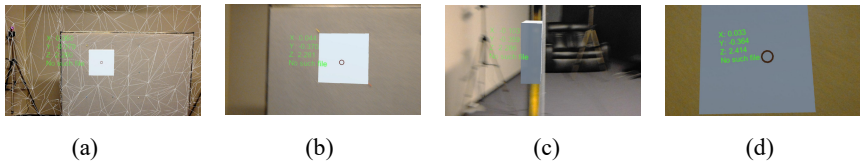

(a)  (b)  (c)  (d)

Figure 6. Procedure for and results of the spatial mapping experiment. (a) The spatial mapping process. (b) Front view of the box attachment results. (c) Side view of the results. (d) Back view of the results. These photographs present the recorded positions of a holographic box used to measure the gap or overlap between the hologram and the target surface to which it was attached.

*2) Materials:* For this experiment, we created an application that allows a holographic box (0.3 m×0.3 m×0.1 m) to be attached to a target surface through spatial mapping, as shown in Figure 6a.

*3) Procedure:* In this experiment, first, the participants were asked to look around to scan the target surface. Then, they were asked to record the positions of the markers on the target surface and attach the holographic box to those positions from different distances (0.5 m to 3.5 m with a step size of 0.5 m). Once the holographic box had been mapped onto the surface, as the third step, the positions of the holographic box were recorded from three views to calculate the gap or overlap between it and the target surface. The experimental results are shown in Figure 6b - 6d, presenting the front, side and back views, respectively, of the box attachment results achieved through spatial mapping. The red circle is the cursor on which the user could focus, and the green text shows the position information of the cursor.

TABLE 1. ACCURACY DEVIATIONS FOR SPATIAL MAPPING, $\Sigma_A^S$

| Distance (m) * | 0.5 | 1 | 1.5 | 2 | 2.5 | 3 | 3.5 |
|---|---|---|---|---|---|---|---|
| Accuracy Deviation $\sigma_A^S$ (%) | 76 | 72 | 71 | 77 | 70 | 74 | 77 |
| * The distance between the HoloLens unit and the target surface ||||||||

*4) Results:* The accuracy deviations $\sigma_A^S$ of the spatial mapping results are shown in Table I. The average $\sigma_A^S$ value is 73.8%, with a standard deviation of only 2.70%. Since the distance is calculated between the hologram and the target surface, the results depend only on the spatial mapping algorithm and are not influenced by possible reconstruction problems. Although the deviations for different distances are only slightly different, the best spatial mapping results are achieved at 1.5 m and 2.5 m, which may be beneficial for high-accuracy tasks such as mapping mechanical components.

## Spatial Mapping

*1) Design:* The hologram visualization experiment was per- formed to evaluate the visual perception effect for HoloLens users. Since the holographic models built in HoloLens applications are created using metric values, users can visualize holograms of the same scale as real objects through the optical lenses. In this experiment, the hologram visualization performance was evaluated by calculating the accuracy deviation $\sigma_A^V$ (in a manner similar to Equation 1) representing the visual deviation between a visualized hologram and a corresponding real object in recorded photographs.

*2) Materials:* To evaluate the hologram visualization performance, we created a 0.25 m×0.2 m×0.1 m holographic box in a self-developed HoloLens application and built a real box of the same size, both of which are shown in Figure 7a.

*3) Procedure:* In this experiment, the participants were asked to first move the visualized hologram to overlap with the real box as exactly as possible and then to take three photographs of the overlapped boxes from the front, side and top views. These three photographs were used to evaluate the visual deviations of the visualized hologram.

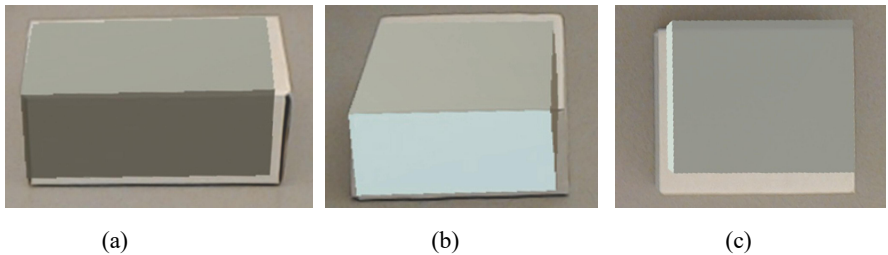

(a)          (b)          (c)

Figure 7. Three views of the overlapping effect between a real object and a corresponding hologram. (a) Front view. (b) Side view. (c) Top view. These recorded photographs were used to calculate the visual deviations between the visualized hologram and the real object.

*4) Results:* An example of results acquired from all three views is shown in Figure 7, where the gray box is the visualized hologram and the white box is the real box used for testing. It is shown that the visualized hologram exhibits good overlap with the real box. The hologram has a slight shift in the top view, which is due to the head posture approximation error of HoloLens when the user moves his/her head quickly. The accuracy deviations of the length, width and height collected throughout the entire experiment are shown in Figure 8, where the error bars represent the standard deviation for each condition. The average $\sigma_A^V$ value is 6.64%, with a standard deviation of 3.29%, which proves that holograms can be visualized precisely using HoloLens. The average difference in terms of the Euler distance is 1.25 cm, with a standard deviation of 0.25 cm. To the human eye, the hologram is essentially the same size as the real object.

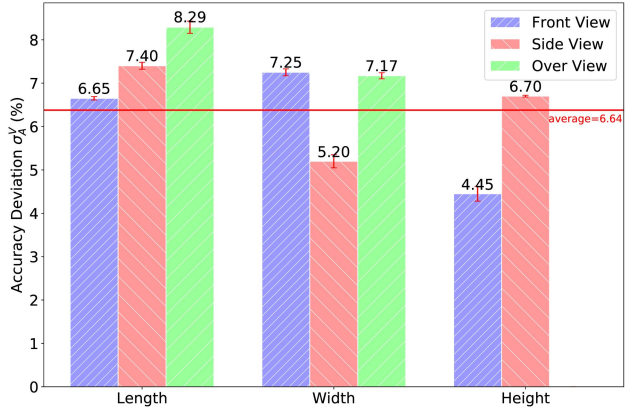

Figure 8. The accuracy deviations $\sigma_A^V$ for hologram visualization. The size of the visualized box was 0.25 m×0.2 m×0.1 m. Therefore, the average deviation in terms of the Euler distance is 1.25 cm, with a standard deviation of 0.25 cm.

## Speech Recognition

*1) Design:* The speech recognition experiment was conducted to evaluate the reliability of controlling HoloLens using voice commands. We tested both system-defined and user-defined commands to obtain an overall evaluation. The user-defined commands (typed into the measurement application) were selected from Wobbrock's paper ("move", "rotate", "delete", "zoom in/out", "open", "duplicate", "previous" and "help"),[10] and several system-defined commands are also chosen ("select", "place", "face me", "bigger/smaller", "ad- just", "remove", "Hey Cortana, shut down" and "Hey Cortana, take a picture").

*2) Procedure:* In this experiment, first, the participants were asked to practice speaking the 8 user-defined commands and the 8 system-defined commands. Each command was required to be identified by HoloLens at least 5 times. Then, the participants were asked to speak each command 10 times in a random order. The number of recognized commands was counted to evaluate the speech recognition capability of HoloLens. All participants had been living in an English- speaking environment for more than 2 years and could speak English fluently. The native languages of the participants were English (6), Chinese (7), Arabic (3), French (2) and Hindi (1).

To analyze the speech recognition capability of HoloLens, we computed the agreement rates $A_r$ for the selected commands.[10-11] The agreement rate $A_r$ represents the level of consensus among the participants for a specific referent $r$ and is defined as

$$A_r = \sum_{P_i \subset P_r} \left(\frac{|P_i|}{|P_r|}\right)^2 \qquad (2)$$

where $P_r$ is the set of operation commands for referent $r$ and $P_i$ is a subset of $P_r$. The value of $A_r$ ranges from $|P_r|^{-1}$ to 1, where $|P_r|^{-1}$ indicates no agreement and 1 indicates perfect agreement.

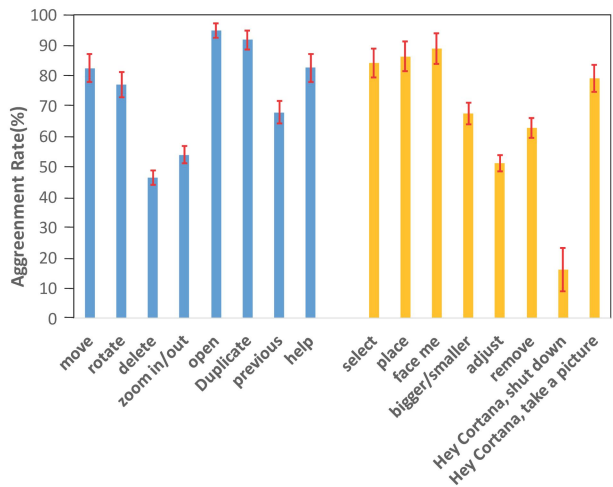

Figure 9. The agreement rates $A_r$ for speech recognition, where the blue and yellow bars represent the agreement rates $A_r$ for the user-defined commands and system-defined commands, respectively.

*3) Results:* The agreement rates for each selected referent are shown in Figure 9, where the blue and yellow bars represent the agreement rates $A_r$ for the user-defined commands and system-defined commands, respectively, and the error bars represent the standard deviation for each condition. The average agreement rates $A_r$ for the user-defined commands and the system-defined commands are 74.47% and 66.87%, respectively, with standard deviations of 16.21% and 22.77%, respectively. A system malfunction occurred during the testing process: the system-defined command "Hey Cortana, shut down" could be correctly recognized but could not call the relevant event. In addition, during the testing process, it was difficult for HoloLens to recognize phrases when a participant spoke two words separated by only a very short pause or no pause.

# DISCUSSION

In this section, we summarize our experiments, discuss about their values, and also present their limitations.

## Summary of Experiments

From the experimental results presented above, we can see that HoloLens is able to estimate the user's head posture more properly at low movement speeds, reconstruct the environment most precisely for flat surfaces under bright conditions, anchor augmented contents at desired locations most accurately at distances of 1.5 m and 2.5 m, display objects with an average size error of 6.64%, and recognize speech commands with correctness rates of 74.47% and 66.87% for user-defined and system-defined commands, respectively.

Our evaluation results can be useful references for future studies. For instance, if a person were to develop a HoloLens shooting game, she/he should be aware that displaying two targets separated by a long distance within a short time should be avoided because this may result in head localization failure, as seen from our first experiment.

## Limitations of Experiments

Although our experiments covered all functional components of the HoloLens system, each of the experiments focused only on one or two specific technical aspects instead of the overall performance. For example, the user control component also offers functionalities for gesture commands and remote control, which were not evaluated. In addition, there are some limitations related to the experimental design. The number of participants was limited, and all participants were university students and employees; moreover, the capabilities of HoloLens were tested only in a limited environment (room, lighting conditions, objects, etc.). Consequently, the results could be different for other groups and other conditions. The actions of the participants during the experiments could also introduce bias.

## CONCLUSION

In this paper, we first introduced the operational mechanism of the HoloLens AR device and then technically evaluated all of its functional components based on related experiments. Experimental results were discussed in details to evaluate HoloLens' capabilities. Limitations of our experiment designs were also illustrated. Our work is expected to be of value for future research and development in this field.

In the future, we plan to evaluate additional performance aspects of HoloLens and use the obtained experimental results as references when designing HoloLens applications.

## ABOUT THE AUTHORS

**Yang Liu** is currently pursuing his M.A.Sc. degree at the Multimedia Computing Research Laboratory (MCRLab), University of Ottawa. He is a student member of IEEE. His research interests include computer graphic and computer vision. He received his B.Eng. degree in information engineering from Beijing Technology and Business University (China). Contact him at yliu344@uottawa.ca.

**Haiwei Dong** is a Research Scientist at the School of Electrical Engineering and Computer Science at the University of Ottawa. His research interests include robotics, multimedia, and artificial intelligence. Dong received his Ph.D. from the Graduate School of Engineering at Kobe University. He is a Senior Member of IEEE and a licensed Professional Engineer. Contact him at hdong@uottawa.ca.

**Longyu Zhang** the corresponding author, is currently pursuing his Ph.D degree in Electrical and Computer Engineering at Multimedia Communications Research Laboratory (MCRLab), University of Ottawa. He has been awarded Canada NSERC Postgraduate Scholarship (Doctoral) for 2015-2018. His research interests include augmented reality, computer vision, and multimedia. Contact him at lzhan121@uottawa.ca.

**Abdulmotaleb El Saddik** is a Distinguished University Professor at the University of Ottawa. His research focus is on multimodal interactions with sensory information in smart cities. He is an ACM Distinguished Scientist, a Fellow of the Engineering Institute of Canada, a Fellow of the Canadian Academy of Engineers and an IEEE Fellow. He received the IEEE I&M Technical Achievement Award and the IEEE Canada Computer Medal. Contact him at elsaddik@uottawa.ca.